# Nanomechanics of Antimonene Allotropes

Tanmay Sarkar Akash[1], Rafsan A.S.I. Subad[2], Pritom Bose[1*], Md Mahbubul Islam[3]

[1]Department of Mechanical Engineering, Bangladesh University of Engineering and Technology, Dhaka-1000, Bangladesh

[2]Department of Mechanical Engineering, University of Massachusetts Dartmouth, 285 Old Westport Road, Dartmouth, MA 02747-2300

[3]Department of Mechanical Engineering, Wayne State University, 5050 Anthony Wayne Drive, Detroit, MI, 48202, USA

*E-mail : bose.buet@gmail.com

**ABSTRACT**

Monolayer antimonene has drawn the attention of research communities due to its promising physical properties. But mechanical properties of antimonene is still largely unexplored. In this work, we investigate the mechanical properties and fracture mechanisms of two stable phases of monolayer antimonene- the α antimonene (α-Sb) and the β antimonene (β-Sb), through molecular dynamics (MD) simulations. Our simulations reveal that stronger chiral effect results in a greater anisotropic elastic behavior in β-antimonene than in α-antimonene. In this paper we focus on crack-tip stress distribution using local volume averaged virial stress definition and derive the fracture toughness from the crack-line stress. Our calculated crack tip stress distribution ensures the applicability of linear elastic fracture mechanics (LEFM) for cracked antimonene allotropes with considerable accuracy up to a pristine structure. We evaluate the effect of temperature, strain rate, crack-length and point-defect concentration on the strength and elastic properties. Tensile strength goes through significant degradation with the increment of temperature, crack length and defect percentage. Elastic modulus is less susceptible to temperature variation but is largely affected by the defect concentration. Strain rate induces a power law relation between strength and fracture strain. Finally, we discuss the fracture mechanisms in the light of crack propagation and establish the links between the fracture mechanism and the observed anisotropic properties.

# 1. INTRODUCTION

2D materials have attained wide attention in this era of nano-material research due to their unique material properties. Since the advent of graphene, researchers have progressed a long way to predict and synthesize other two dimensional materials for instance, Group IVA (tetragens) elemental analogues like silicence,[1] germenene,[2] hexagonal boron nitride (h-BN);[3] Group VA (pnictogens) elements such as phosphorene,[4] arsenene,[5] antimonene,[6] bismuthene;[7] monolayer TMDs like $MoS_2$,[8], $MoSe_2$,[9] $WS_2$,[10] $TiS_2$,[11] $InSe$[12] etc. Even though being mechanically super elastic[13] and electrically super conductive,[14] graphene flounders when it comes to the application where a certain amount of band gap is mandatory. TMDs and pnictogens dominate and possess further prospects in regard of this special requirement in many electrical and optoelectronic nano-devices.[15–18] As soon as the 2D form of phosphorus (black phosphorus/ phosphorene) emerged as Filed Effect Transistor,[19] it took a very little interval to verify that all other Group puckered VA elements also have stable, freestanding 2D structures (entitled as nitrogene, phosphorene, arsenene, antimonene, and bismuthene). Among these pnictogens, antimonene allured particular prominence because of its some distinct features such as, fundamental band gap in semiconducting antimonene monolayer 0.3-2.28 eV[20–22] which can be tuned by varying its chirality, width,[23] and applying tensile strain,[24] high stability in ambient conditions[25] and almost no signs of deterioration over months[26] which is the blatant disparity to its predecessor – phosphorene, capability of its single layer to produce binary compound[27] that demonstrates topologically non-trivial characteristics under definite settings, competence of absorbing a wide range of wave length and high carrier mobility,[21] and so on. These outstanding qualities of 2D antimonene makes it suitable for its utilization in copious applications for instance, field effect transistors (MOSFETs),[28] optical nano-devices,[16] photo electric devices,[29] high efficiency quantum computation[30] as topological insulator[31,32] etc.

Antimonene monolayer has already been fruitfully synthesized from bulk antimony by mechanical isolation of few-layer antimonene flakes,[26] epitaxial growth on $PdTe_2$,[25] on Ge substare,[33] liquid phase exfoliation.[34] Remarkably, flat antomonene has also been synthesized very recently on Ag substrate.[35]

However, there are plenty of works regarding the electrical, optical and magnetic properties of monolayer antimonene but a very few of on mechanical.[23,36,37] So, to inquest further on the mechanical characteristics of SLSb is wanting. There are also studies that report modulating not only electronic but also magnetic properties inducing defect and tensile strain which eventually resemble the necessity of finding out the mechanical properties and fracture mechanism of SLSb nano-sheet at atomic scale.[38–40] Although in the perspective of theoretical inquiry first principle study may reveal the basic structural characteristics of a nano-material incorporating with a few sets of atom of that material,[41] MD investigation is a must in order to have an insight into the fracture mechanism of a nano-sheet meticulously while dealing with a lot. And, to our best knowledge, there is hardly any study of MD on SLSb to explore its exhaustive mechanical properties.

In this work, we investigated the mechanical characteristics of monolayer Sb sheet for its two different structures (α, and β) varying temperature, strain rate, crack-length and defect concentration ranging from 1K to 500K, $10^8$ $s^{-1}$ to $10^{10}$ $s^{-1}$, 0 to 110 Å and 0% to 5% accordingly. This paper focuses on calculating the stress distribution at critical point, eg., crack tip, using local volume averaged virial stress. The derived stress distribution is used to determine the applicability of LEFM at nanoscale and also used to derive the fracture toughness at different crack lengths. Although the strain rate range is quite higher compared to the real life scenario, it is frequent in MD simulations to qualitatively assess the impact of strain rate at this order.[42] While we changed the temperature, we preserved the strain rate constant ($10^9$ $s^{-1}$), and vice-versa for fixed temperature (300K). And we kept the both temperature (300K), and strain rate ($10^9$ $s^{-1}$) constant while altering the defect concentration. Finally, we demonstrated the fracture mechanism of pristine and defected Sb nano-sheet for all three structures considered here in details.

## 2. METHODOLOGY

Monolayer antimonene nano-sheets (20nm x 20nm) for both α-Sb and β-Sb are generated with MATLAB[43] script and Atomsk[44] codes. The lattice parameters for Sb are used as follows- lattice constant and bond length of α-Sb: 4.12 Å and 2.89 Å respectively; $a_1$, $a_2$, $a_3$ for β-Sb: 4.73 Å, 4.36 Å, and 11.11 Å respectively.[45] In order to study the influence of defect concentration on the mechanical behavior of SLSb, sheets having 1% to 5% of point defects are generated by deleting atoms from random sites. Four sheets with different defect arrangements are simulated for each defect concentration to derive statistically sound results. We apply gradually increasing uniaxial tensile strains on the antimonene sheet and record the corresponding virial stress responses. OVITO software[46] is used to visualize the fracture mechanism of monolayer Sb. LAMMPS[47] simulation package is employed to execute all the simulations. We preserve periodic boundary condition in the lateral directions (X and Y) to elude the finite size effect. Non-periodic boundary condition is maintained in the out of plane direction to avoid interaction between periodic images. Conjugate gradient (CG) scheme is carried out for energy minimization. Then the structure is relaxed for one 100 pico-seconds (ps) within an isobaric-isothermal (NPT) ensemble with slow damping of pressure (0.05 ps) and temperature (0.5 ps). A time step of 1 fs is used to ensure proper convergence. Stress-strain relationship is developed by straining the simulation box uni-axially and computing the average stress over the structure. Virial stress is taken as the basis of estimating the average. Relation among virial stress components are as follows:

$$\sigma_{virial} = \frac{1}{\Omega} \sum_i \left( -m_i \dot{u}_i \otimes \dot{u}_i + \frac{1}{2} \sum_{j \neq i} r_{ij} \otimes f_{ij} \right) \quad (1)$$

where the summation over all the atoms occupying the total volume is denoted by $\Omega$, $\otimes$ specifies the cross product, the mass of atom i is represented by $m_i$, $r_{ij}$ is the position vector of atom, the displacement of an atom with respect to a reference point is directed by time derivative $\dot{u}_i$, and $f_{ij}$ is the interatomic force exerted on atom i by atom j. Engineering strain is evaluated considering the following equation:

$$\varepsilon = \frac{l - l_0}{l_0} \quad (2)$$

$l_0$ is the undeformed length of the box and $l$ is the instantaneous length.

We utilized a recently developed Stillinger–Weber (SW) potential by *Jiang et al.*[45] to address the interatomic interactions in this study. The SW potential includes a two body term and a three body term characterizing the bond stretching and bond breaking, respectively. The mathematical relations are as follows:

$$\Phi = \sum_{i<j} V_2 + \sum_{i>j<k} V_3 \tag{3}$$

$$V_2 = A e^{\left[\frac{\rho}{r-r_{\max}}\right]} \left(\frac{B}{r^4} - 1\right), \tag{4}$$

$$V_3 = K \theta e^{\left[\frac{r_1}{r_{ij}-r_{\max\ ij}} + \frac{r_2}{r_{ik}-r_{\max\ ik}}\right]} (\cos q - \cos q_0)^2 \tag{5}$$

Here, $V_2$ is the two body bond stretching and $V_3$ is the angle bending terms. The terms $r_{\max}$, $r_{\max\ ij}$, $r_{\max\ ik}$ are the cutoffs and the angle between two bonds at equilibrium configuration is symbolized by $\theta_0$. $A$ and $K$ imply energy related parameters that are established on Valance Force Field (VFF) model. $B$, $\rho$, $\rho_1$, and $\rho_2$ are other parameters that are fitted coefficients. These parameters and their corresponding value can be found in ref. [45]. For the current SW potential parameters, we found a negative out of plane Poisson's ratio at strain larger than 15% (Figure S1).

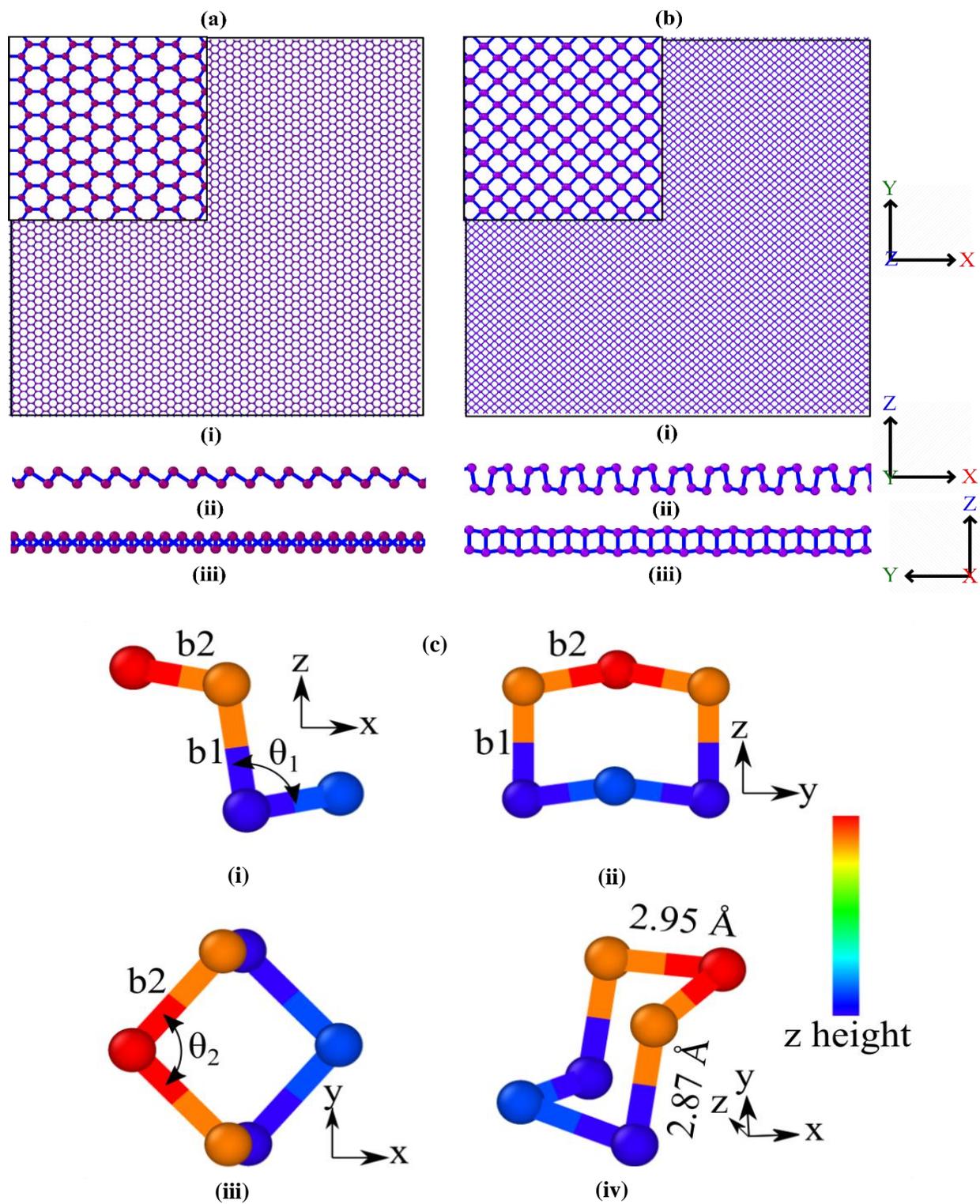

Figure 1: Atomic structure of monolayer (a) α, and (b) β antimonene nanosheet; (i), (ii), and (iii) represent top, front and side view accordingly. (c) structural parameters of β structure shown in (i)front (ii)side (iii)top and (iv)top views

# 3. METHOD VALIDATION

Stress-strain graph for temperature 1K of α-Sb and β-Sb pristine are presented in Figure 2. For validation purpose, Table 1 is constructed to compare their intrinsic mechanical properties with literature . From the comparison, it is evident that the results predicted in this study are in well-agreement with the existing literature[45].

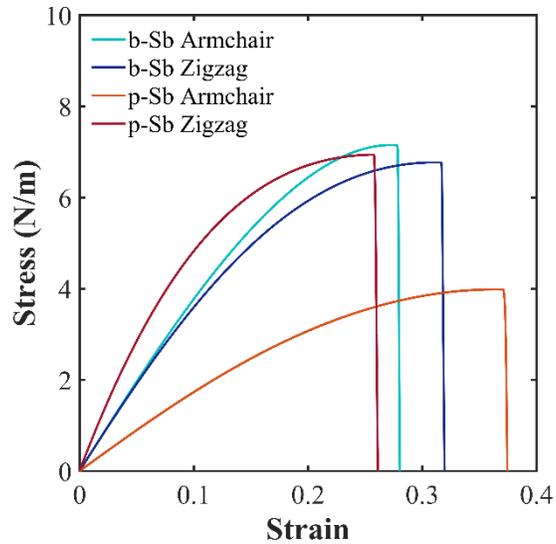

Figure 2: Stress-strain relationship of single layer α-Sb and β-Sb nano-sheets at 1K temperature

Table 1- Validation of our calculation with the existing literature

| Mechanical Properties @1K | | This Study | | Previous Study[45] | |
|---|---|---|---|---|---|
| | | Armchair loading | Zigzag loading | Armchair loading | Zigzag loading |
| Young's Modulus (GPa) | α-Sb | ~39.7 | ~38.7 | ~39.6 | ~39.6 |
| | β-Sb | 18.0 | 58.8 | ~18.3 | ~65.2 |
| Ultimate Tensile Stress (GPa) | α-Sb | 7.1 | 6.7 | ~7.1 | ~6.7 |
| | β-Sb | 3.9 | 6.9 | ~3.7 | ~6.4 |
| Fracture Strain (%) | α-Sb | 28.0 | 31.9 | ~28.0 | ~31.0 |
| | β-Sb | 37.4 | 26.1 | ~37.0 | ~17.0 |

# 4. RESULTS AND DISCUSSIONS

## 4.1 Effect of Temperature

In real-life application, 2D materials might encounter a frequent risk of being exposed to a high temperature due to its particularly smaller dimension. A small current can easily generate a large temperature through joule heating. [48,49] Deterioration of physical properties is often associated with an exposure to high temperature. So, it is essential to inquire the impact of temperature on the material properties. Here, we investigated the influence of temperature on the stress-strain relationships (Figure 3) to determine the elastic properties of two monolayer Sb allotropes. A and β antimonene monolayer sheets are equilibrated in the temperature ranging from 1K to 500K and strained at a constant strain rate of $10^9$ s$^{-1}$ separately along armchair and zigzag directions.

Escalating temperature generally causes a significant worsening of many material parameters (eg., ultimate strength, elastic modulus, and failure strain) regardless of chirality and structural differences. We simulated mechanical properties of α-Sb and β-Sb at different temperatures and the comparison is presented in Figure 4. A small linear decrease of elastic modulus with temperature is observed (Figure 4a(iii), 4b(iii)). Such a material softening is associated with a small thermal expansion arising from the anharmonic contribution of the potential function. Elevated temperature also facilitates the thermal vibrational instabilities. And this unsteadiness assists the possibility of some bonds exceeding the critical bond length and instigating rapid failure. Moreover, the intensification of temperature engenders higher entropy in the material and expedites crack propagation. Material weakening occurs for this reasoning too.[50,51]

We compared the change of elastic modulus, ultimate tensile stress (uts) and fracture strain for α and β structure both for loading along armchair and loading along zigzag direction. α-Sb possesses higher uts and elastic modulus than β-Sb whereas β-Sb shows more softness and larger fracture strain in armchair direction. Nearly perpendicular bond with XY plane in β-antimonene sheet are inclined while it goes under armchair directional tension. Thus, stress does not accumulate as much as develop in the α one and the sheet displays more endurance before failure. Consequently, fracture stress stands less and fracture strain grows higher. In case of zigzag loading, that bond does not extend but creates a hindrance for the zigzag bonds to lengthen. Therefore, stress developed with less effort and the sheet gets harder. Hence, uts and elastic modulus of β-antimonene sheets improves while fracture strain falls relative to α structure this time.

We also measured the change of uts, fracture strain, and elastic modulus of the material with the variations of temperature. In α structure, they decline almost 17.9%, 33.4%, and 6.9% (in armchair) and 13%, 28.6%, and 22% (in zigzag) accordingly. Their changes in β structure are as follows 27.6%, 42%, and 9.6% (in armchair) and 19.5% 38.6% and 4% (in zigzag) for the modification of temperature from 1K to 500K. Measuring and fitting the numerical values, we also suggest a linear relationship of elastic modulus and temperature of the material by the following equation.

For α-antimonene sheet as:

In armchair direction: $y = -0.0054x + 39.395$ 　　　　　　　　　　　　　　　　　　　6

In zigzag direction: $y = -0.0173x + 38.858$ 　　　　　　　　　　　　　　　　　　　7

And for β-antimonene sheet as:

In armchair direction: $y = -0.0034x + 18.065$ 　　　　　　　　　　　　　　　　　　　8

In zigzag direction: $y = -0.0049x + 64.073$ 　　　　　　　　　　　　　　　　　　　9

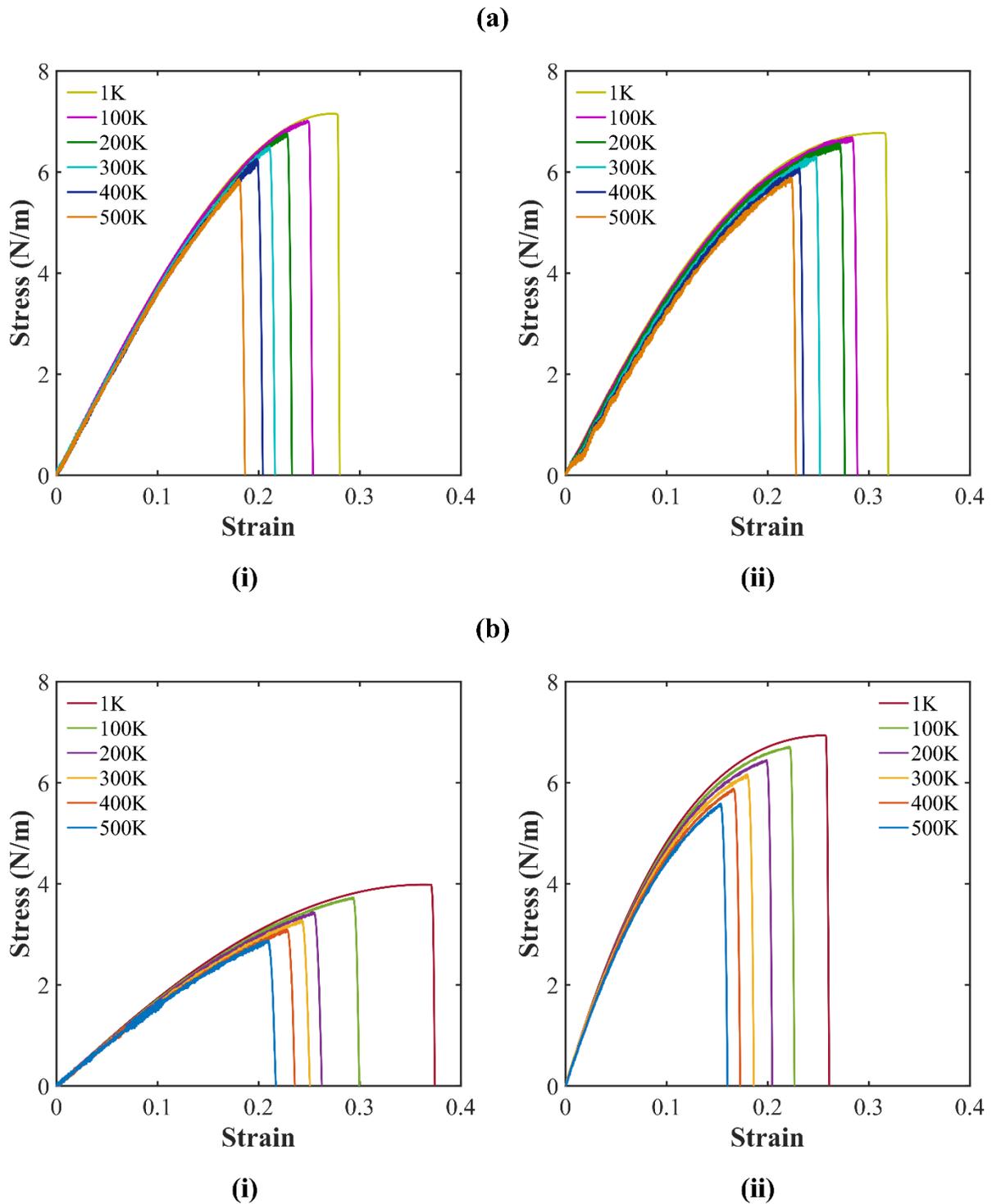

Figure 3: Temperature reliant stress-strain curves for (a) α, and (b) β structure of pristine SLSb sheet; (i), and (ii) denote armchair and zigzag directional loading respectively.

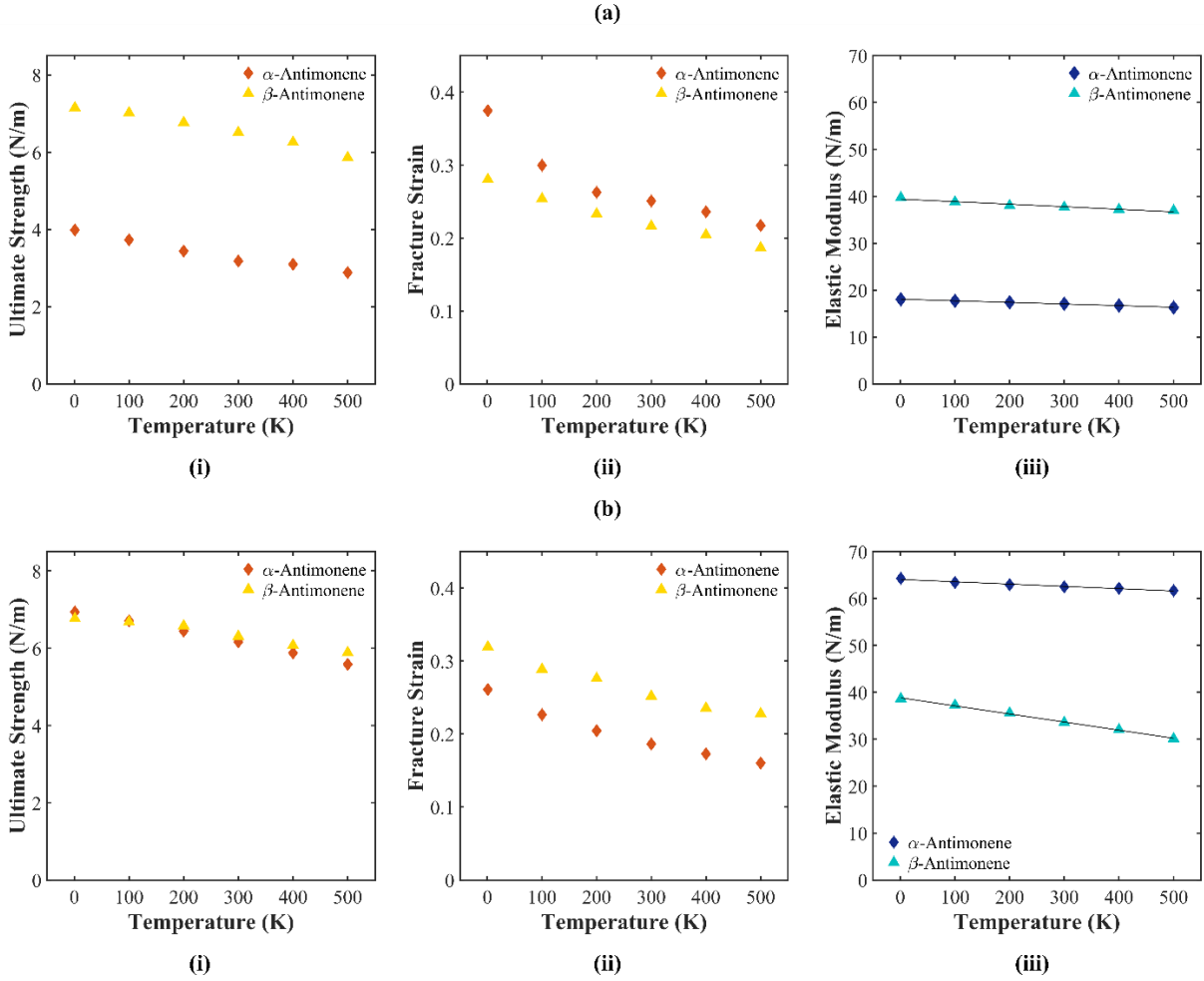

Figure 4: Comparison of (i) ultimate tensile strength, (ii) fracture strain, and (iii) elastic modulus among the planner, α, and β structured antimonene nanosheet with the alteration of temperature; (a), and (b) signify armchair and zigzag directional tension accordingly.

### 4.2 Effect of Strain Rate

Our MD simulations are performed at a strain rate of ~$10^9$ s$^{-1}$ which is few order of magnitude larger than the experimentally practiced strain rate. To demonstrate the strain sensitivity of the mechanical properties, we performed uniaxial tension simulations on the monolayer Sb nanosheet at varying strain rate ranging from $10^8$ s$^{-1}$ to $10^{10}$ s$^{-1}$ (Figure 5). SLSb sheet exhibits less sensitivity for altering strain rate than that for temperature and defect concentration. While the Young's modulus of the material remains virtually unaffected, ultimate tensile strength of armchair and zigzag tension increases about 1.5%, 3.7% for α-antimonene and 5.9%, 4.9% β-antimonene accordingly for shifting the strain rate from $10^8$ s$^{-1}$ to $10^{10}$ s$^{-1}$. That implies the greater the strain rate, higher the strength. Since at higher strain rate, the time needed for the material to respond and relax is less. Hence, it does not permit atomic thermal fluctuations to pass over the energy barrier to break their bonds and cannot foster rearranging bonds, growing vacancy, and propagating crack. Consequently, higher strain rate gives rise to the fracture stress and vice versa.[52–54] These factors are also persistent for SLSb sheet. Figure 6 illustrates higher fracture stress for higher strain rate in logarithmic scale.

We evaluated the sensitivity of ultimate stress on the strain rate by the following equation:[52]

$$\sigma = C\dot{\varepsilon}^m \qquad \qquad 10$$

Here, $\dot{\varepsilon}$ is strain rate, $C$ is a constant, $\sigma$ and $m$ indicate the fracture strength and the strain-rate sensitivity respectively. According to the logarithmic scale the equation can be written as subsequently:

$$\ln(\sigma) = \ln(C) + m\ln(\dot{\varepsilon}) \qquad \qquad 11$$

Estimating (*m*) the slope from the linear fitted data in Figure 6, we propose the equation for monolayer α-antimonene sheet as:

In armchair direction: $y = 0.0041x + 1.7892$ \qquad 12

In zigzag direction: $y = 0.0075x + 1.6866$ \qquad 13

And for β-antimonene sheet as:

In armchair direction: $y = 0.0138x + 0.8980$ \qquad 14

In zigzag direction: $y = 0.0122x + 1.5727$ \qquad 15

Figure 5: Strain rate dependent stress-strain relationships for (a) α, and (b) β structure of single layer Sb nanosheet; (i), and (ii) indicate the loading along armchair and zigzag directions respectively.

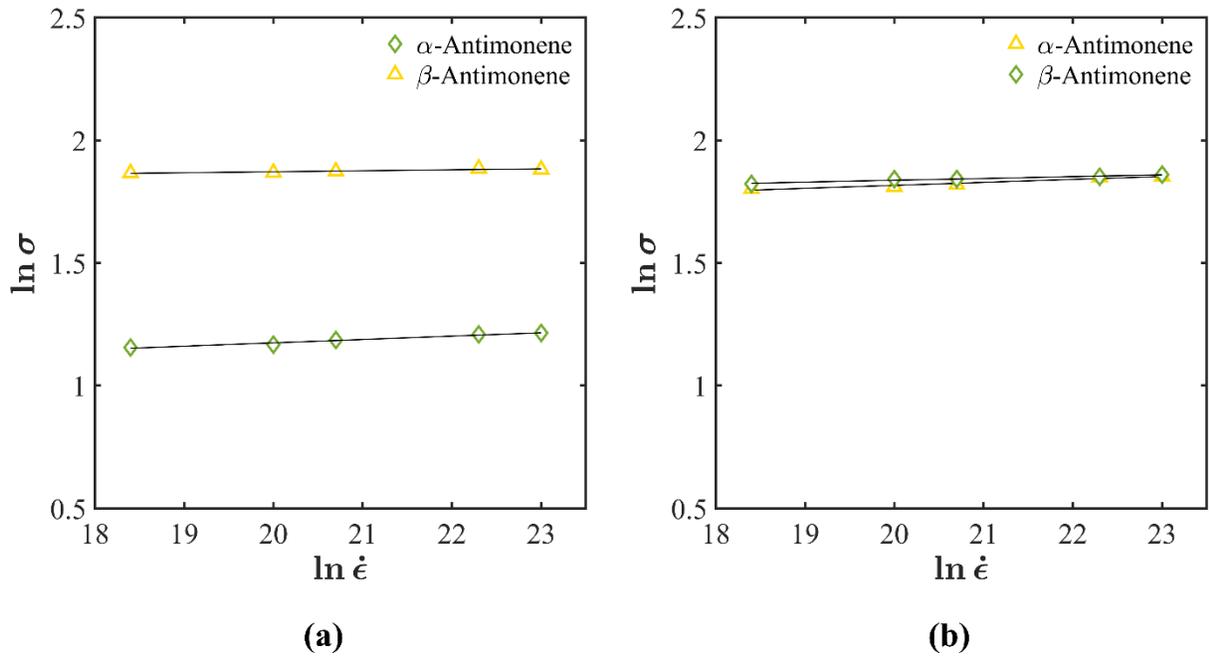

Figure 6: Variation of fracture strength of SLSb nanosheet for (a) armchair (b) zigzag directional tension with respect to strain rate in logarithmic scale

## 4.3 Mechanical Properties of Cracked Sample

### 4.3.1 Crack-tip Stress-field

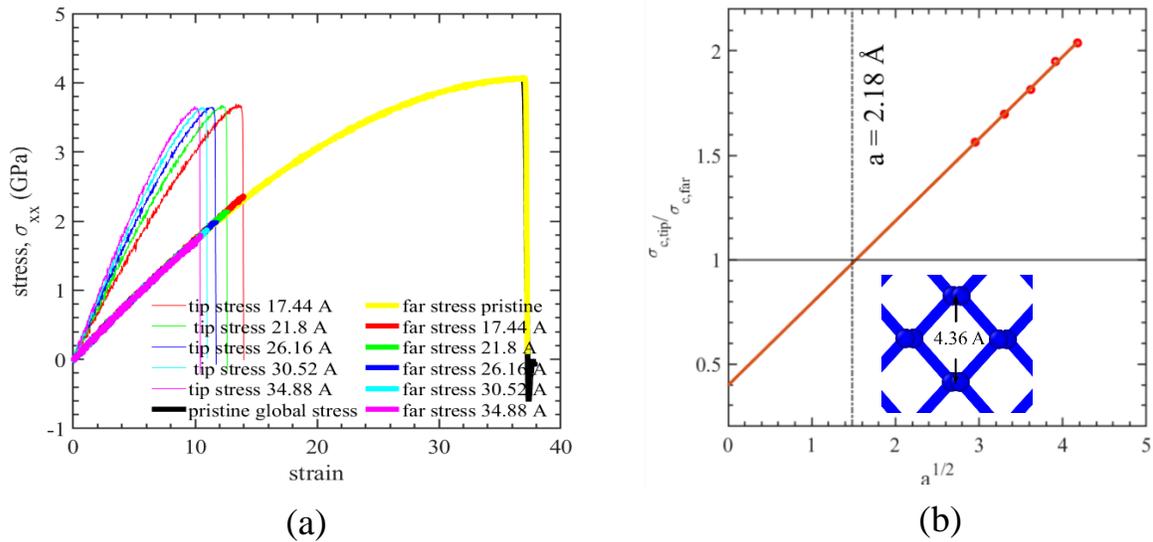

Figure 7: (a) Tip stress and far-field stress for different crack lengths of β-AC Sb (b) Relationship between critical stress concentration factor and $\sqrt{a}$. The relationship predicts a unit stress concentration for pristine structure.

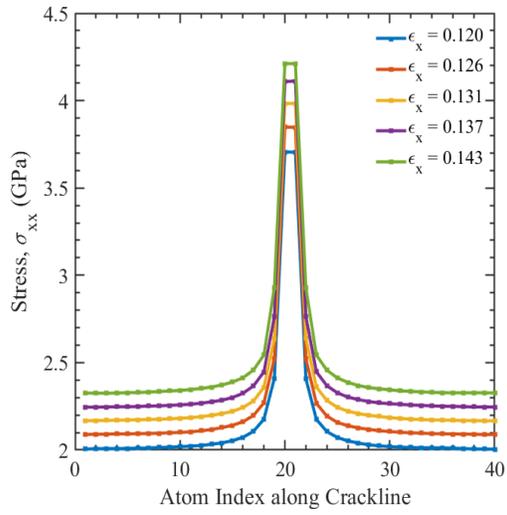
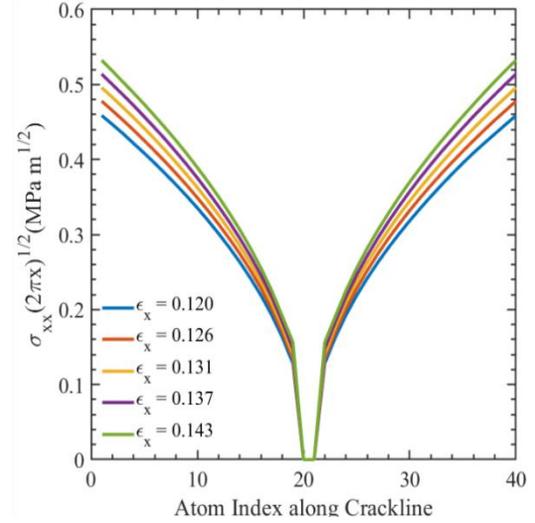

(a)          (b)

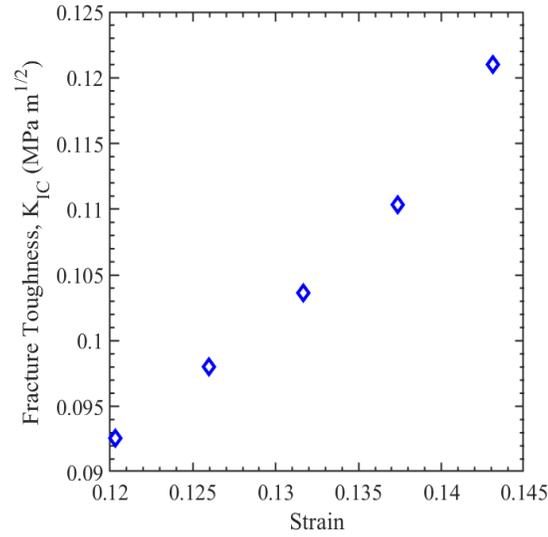

(c)

Figure 8: (a) Crack-line stress evolution (b) Evolution of crack opening stress distribution, $\sigma_{yy}\sqrt{2\pi x}$ (c) Evolution of stress intensity factor $\lim_{x \to 0} \sigma_{yy}\sqrt{2\pi x}$ with applied strain for a fixed crack length of β-AC Sb.

Vacancies and extended defects, viz, cracks act as stress raisers. Here we discuss the stress field generated at the crack tip in the light of molecular dynamics and continuum mechanical models. For brittle materials like antimonene, the stress is almost entirely concentrated to the tip atoms. Thus, a very small fracture process zone is perceived. Near tip solution of continuum fracture mechanics gives the following relation -

$$\sigma_{yy} = \sigma_0 \sqrt{a} \left( \frac{1}{\sqrt{2r}} \right) \hspace{4cm} 16$$

Here, $\sigma_{yy}$ is the normal stress at a 'r' distance from the crack-tip for a geometry with '2a' crack length. Equation 16 predicts an infinite stress at the crack tip (r=0). It happens due to an infinitely sharp crack-tip consideration. However, in practical case, we have a finite curvature at the crack tip and thus a finite stress level is perceived at all applies strains. Figure 7(a) depicts the local virial stress near crack tip and at a far field for different crack lengths of a representative configuration (here, AC β-Sb). The figure confirms that, irrespective of crack lengths, the crack-tip atomic bonds sustain up to a stress level very close to the ultimate strength, i.e., material failure is primarily dictated by the ultimate strength, however, due to the stress concentration, the actual global stress at fracture is only a fraction of the ultimate strength. We plot the stress concentration ($\frac{\sigma_{yy,tip,c}}{\sigma_{0,c}}$) factor at crack tip as a function of the square root half crack length (Figure 7(b)). The stress concentration factor maintains a linear relationship with $\sqrt{a}$, supporting the continuum linear elastic fracture mechanics solution. More interestingly, we find that, this linear relationship predicts a unit stress concentration at a crack length, 2a ~ 4.36Å, i.e., pristine condition. This confirms that no flaw tolerance is perceived in this nanostructure and the strength of the material can be substantially predicted with linear elastic fracture mechanics (LEFM) at all crack dimensions, without the consideration of plastic deformation at the crack tip.

We note that, in this paper local volume averaged virial stress is calculated from per atom virial and local Voronoi volume with a unit nanometer thickness. Per-atom virial was calculated as described in the original paper.[55] Voronoi volumes are calculated by tracing the Voronoi vertices for each atom from the atomic trajectories at each several time steps (supplementary fig S2). Previously, a Voronoi equivalent volume was used to calculate local virial stress for unit cells with a cubic symmetry.[56] Thompson et. al. [55] showed the equivalence of atom-cell, per-atom and group virial formula. They predicted that per atom virial tensor can effectively represent the contribution of atomic virials to the global virial stress. Figure 7(a) corroborates the above prediction due to an exact overlapping of few atom averaged and globally averaged stress-strain relation for a pristine geometry.

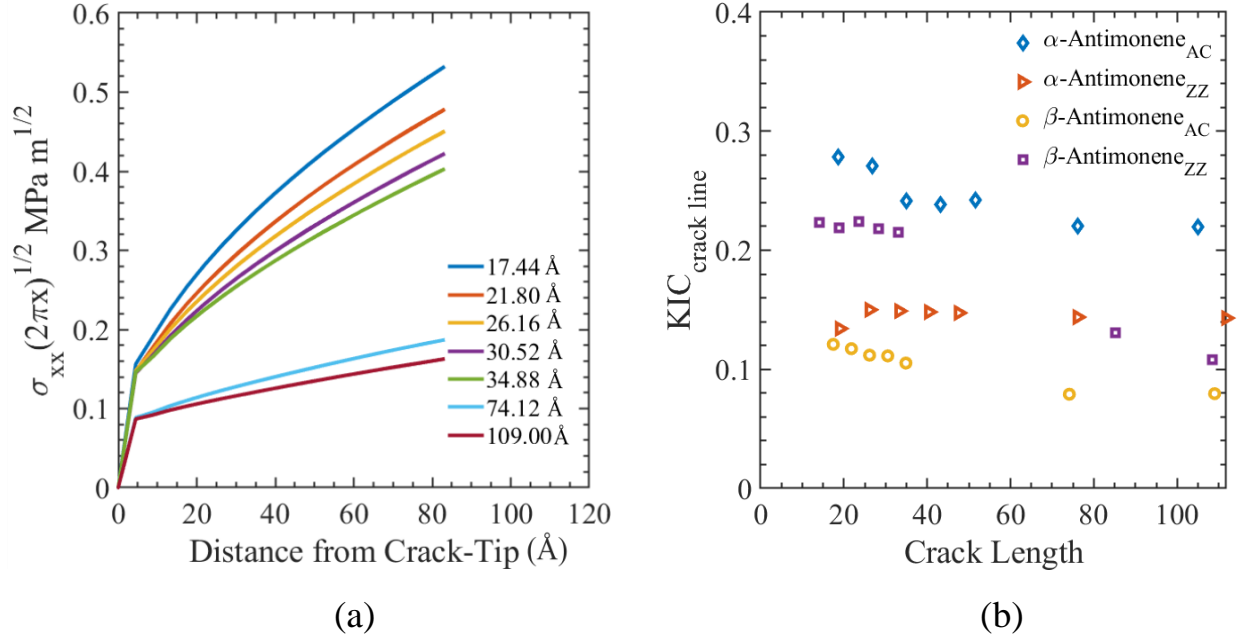

Figure 9: (a) stress intensity factor $\lim_{x \to 0} \sigma_{yy}\sqrt{2\pi x}$ with crack length for a representative configuration (AC β-Sb). (b) Fracture toughness with crack length for different configurations.

Atomically sharp cracks act as stress risers. From figure 8(a), the crack-line normal stress distribution resembles an inverse square root relationship with distance from the crack tip. In such a case, a critical stress intensity factor, i.e., fracture toughness ($K_{IC}$) can be derived by equation 17 that should accurately predict the fracture strength of the material for different crack dimensions.

$$K_I = \lim_{x \to 0} \sigma_{yy}\sqrt{2\pi x} \tag{17}$$

To calculate the fracture toughness with equation 17, we deform the structure in small steps (~0.5%) with intermediate 15 ps relaxations for stress convergence. We average the stress data from the final half of the relaxation steps. This procedure ensures a reliable and smooth crack-line stress distribution $\sigma_{yy}$. Now we calculate the crack opening stress distribution $\sigma_{yy}\sqrt{2\pi x}$ (Figure 8(b)) and linearly extend the curve from the second atomic position to the crack-tip(x=0) which gives the stress intensity factor, $K_I$. Using the same procedure we calculate $K_{IC}$ for the crack-line stress distribution immediately before the crack-tip disruption. Figure 8(c) shows a typical evolution of $K_I$ with strain.

Figure 9(b) plots the fracture toughness of the two orientation of β-Sb and α-Sb for different crack lengths between 15 to 110 Å. The calculated $K_{IC}$ show that for crack lengths above about 60 Å, fracture toughness converges to constant values for different configurations. According to Griffith's analysis, fracture toughness is a material constant. However, nanoscale cracks are prone to show deviations from the Griffith's prediction. We observe fluctuations in $K_{IC}$ at smaller crack lengths. Some authors often attribute this behavior with the crack tip plastic deformation which is specially plausible for materials with doped foreign species that superimpose their own stress field deforming the tip plastically. [57] However, earlier in this section we observed that for this material, crack tip behavior is convincingly described without the need for plastic dissipation. Thus, it is expected that, the observed fluctuation of $K_{IC}$ is due to non-singular part of stress field, i.e., higher order terms in the expansion of stress field that are only dominant for smaller crack lengths.

### 4.4 Effect of Defect Concentration

In practical life, harsh chemical settings during manufacturing process, and recurrent subjection to high temperature make Sb nano-sheet vulnerable to the growth and evolution of defects which eventually minimizes the material property drastically. In many cases, such defects are unavoidable. However, sometimes they are also intentionally placed upon the nanostructures to achieve preferred properties, exclusively in terms of electrical and optical properties and even to enhance mechanical stability for some nanomaterials.[38,58]

To measure the susceptibility of material properties on the density of defects, stress-strain curves for the mentioned structures of SLSb have been constructed and demonstrated in Figure 10. The figure suggests that arbitrarily dispersed and rising quantity of defect density can extensively minimize the material integrity. It is because the vacancy defects always act as fracture initiation points and occupancy of defects at numerous locations of the sheet augments the possibility of nucleation events at those positions.[59] As a result, the material gets fragile.[60,61]

We also perceive similar trend of fracture strength, fracture strain and elastic modulus between armchair and zigzag directional loading of α-antimonene and β-antimonene structures as stated earlier in the effect of temperature section. The Young's modulus of α structure and fracture strain of β in armchair direction; Young's modulus in β and fracture strain of α remains higher. The comparison has been demonstrated in Figure 11. Ultimate strength, fracture strain and Young's modulus of α-Sb and β-Sb nanaosheet diminish nearly 35.7%, 53.2%, 63.8% and 36.2%, 45.2%, 60.8% (in armchair); 40.7%, 54.4%, 60.9% and 27.6%, 54.1%, 56.7% (in zigzag) at 5% defect concentration compared to pristine one. It also ensures that increasing defect density has way much impact in Sb nanosheet in proportion to the influence of temperature and strain rate.

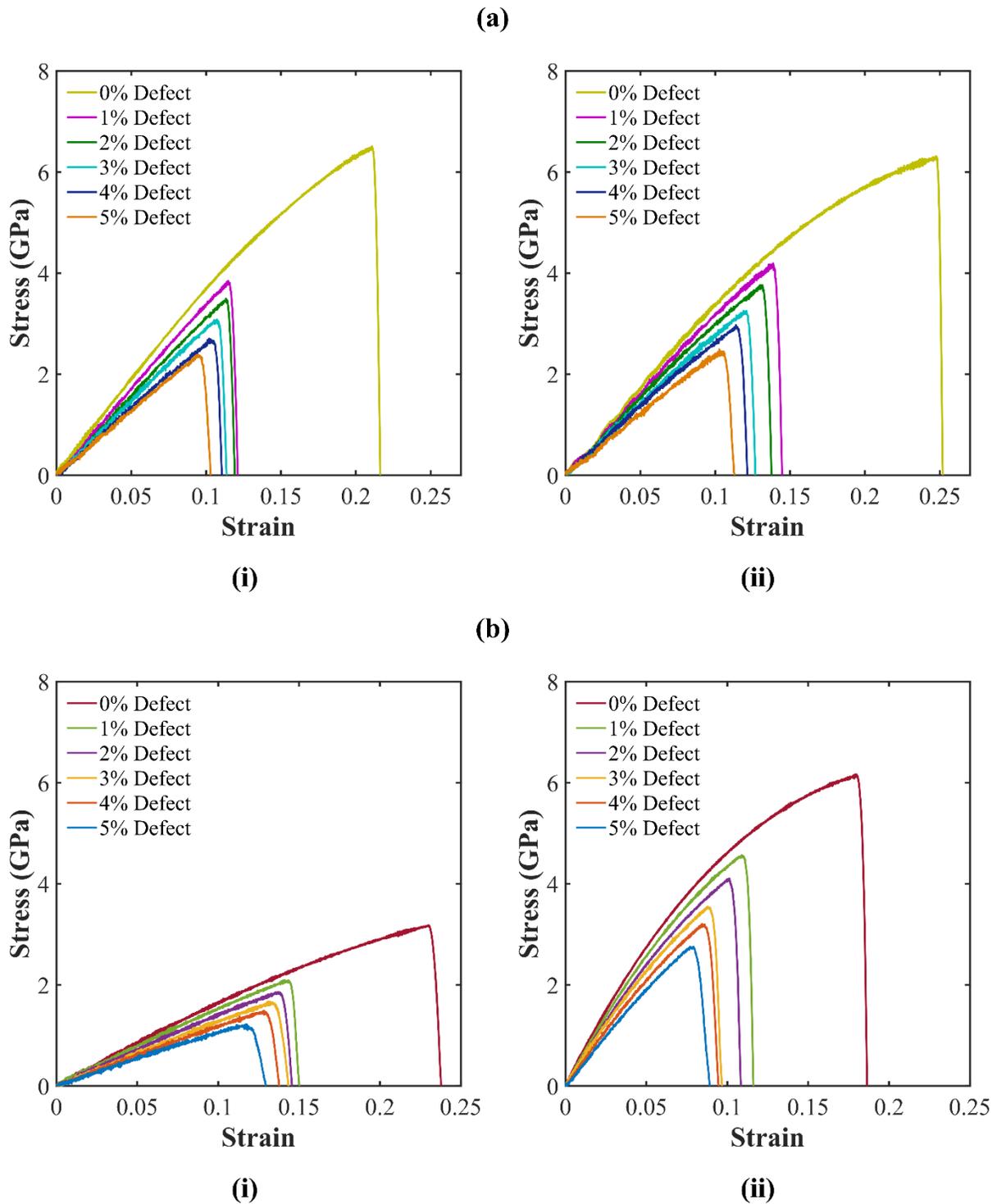

Figure 10: Stress-strain relationships for the defected SLSb of (a) α, and (b) β samples varying the defect concentration while applied strain along (i) armchair, and (ii) zigzag direction

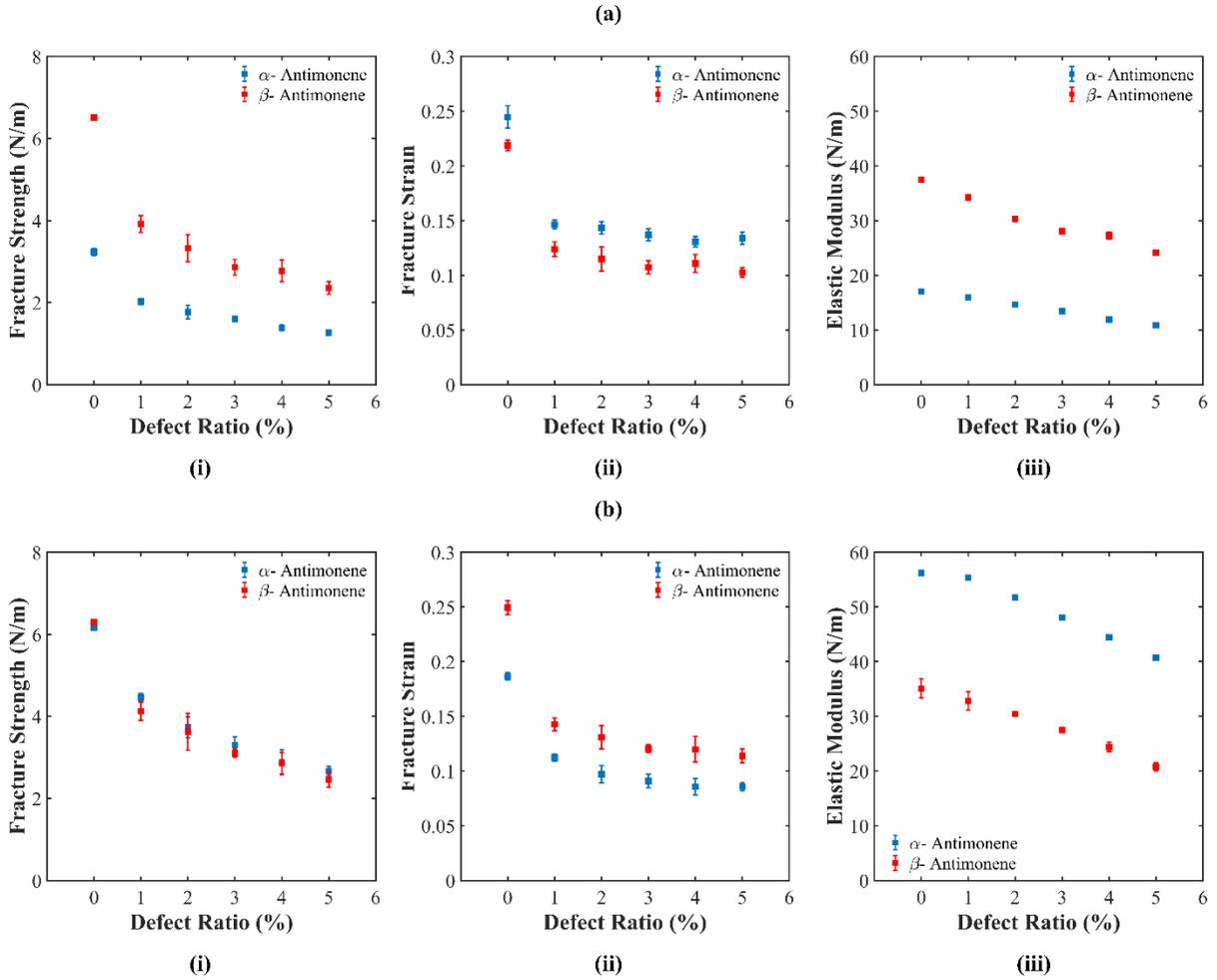

Figure 11: Estimation of (i) fracture strength, (ii) fracture strain, and (iii) elastic modulus for α, and β structures of monolayer Sb sheet with the change of defect density; (a), and (b) connote armchair and zigzag directional tension respectively.

### 4.5 Fracture Mechanism

In order to demonstrate the fracture mechanism of α-antimonene and β-antimonene nanosheet we generated a single vacancy by deleting atom at the center of the sheet and applied tension along both armchair and zigzag directions. The bond breaks at first in case of α-SB, is perpendicular to the loading and gets the opportunity to become almost parallel to the XY plane while undergoing tension. When the crack forms in armchair direction, it faces four (two at each side) evenly inclined bonds situated at the crack tip that offer two potential crack propagation paths (±60° with X direction) (Figure 12). Hence, branching phenomenon takes place during armchair directional crack propagation (Figure 13). Nevertheless, when zigzag crack formulates, it just confronts two bonds at the crack tip which is perpendicular to the crack. Therefore, it does not generate any branch initially (Figure 12).[62] Yet, it might encounter some weaker bonds due to the vibrational instabilities in other directions rather than perpendicular in its long propagation path. In that case branching might occur in the distant path for zigzag directional crack too. The phenomenon is also spotted here and represented in Figure 13. In β Sb structure, there are bonds positioned nearly vertical to the XY plane. Thus, when it experiences armchair (X-axis) directional tension, it tries to descent and ultimately breaks while crosses its critical length. As the vertical bonds are at the same line along Y-axis, the broken bond engenders a straight path for the crack to follow (Figure 12). Thus, almost no branching arises in zigzag directional crack propagation path. Again, tension in zigzag direction (Y-axis), since the zigzag bonds take the exerted load, it does not extend much (Figure 12). Therefore, the breaking occurs at the zigzag directional bonds only and causes branching. The phenomena have been depicted in Figure 10.

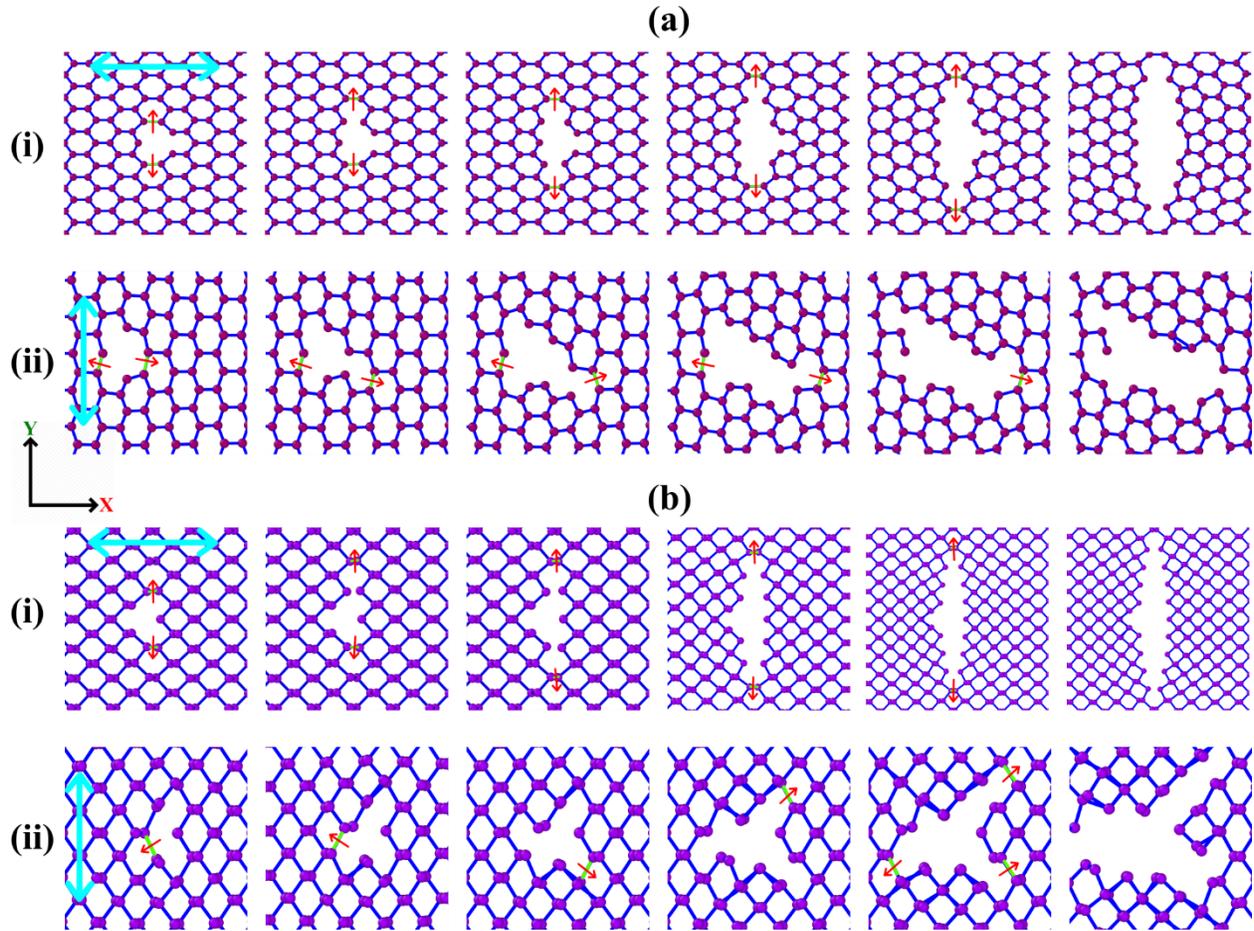

Figure 12: Bond breaking and crack propagation path for (a) α-antimonene, and (b) β-antiomone while straining along (i) armchair, and (ii) zigzag direction represented by X and Y axis respectively. Green colored bonds indicate the bonds to be broken at real time; red color arrows show the crack propagation path and cyan color arrows symbolize the loading directions.

As a whole, since tensile straining is executed, stress starts to build up in the structure. Nonetheless, stress does not develop homogeneously rather mostly around the vacancy positions in the perpendicular direction of the loading. And nucleation occurs at those stress concentrated areas. Following the nucleation, crack forms and propagates perpendicular to loading direction (in zigzag direction for armchair loading and vice versa).

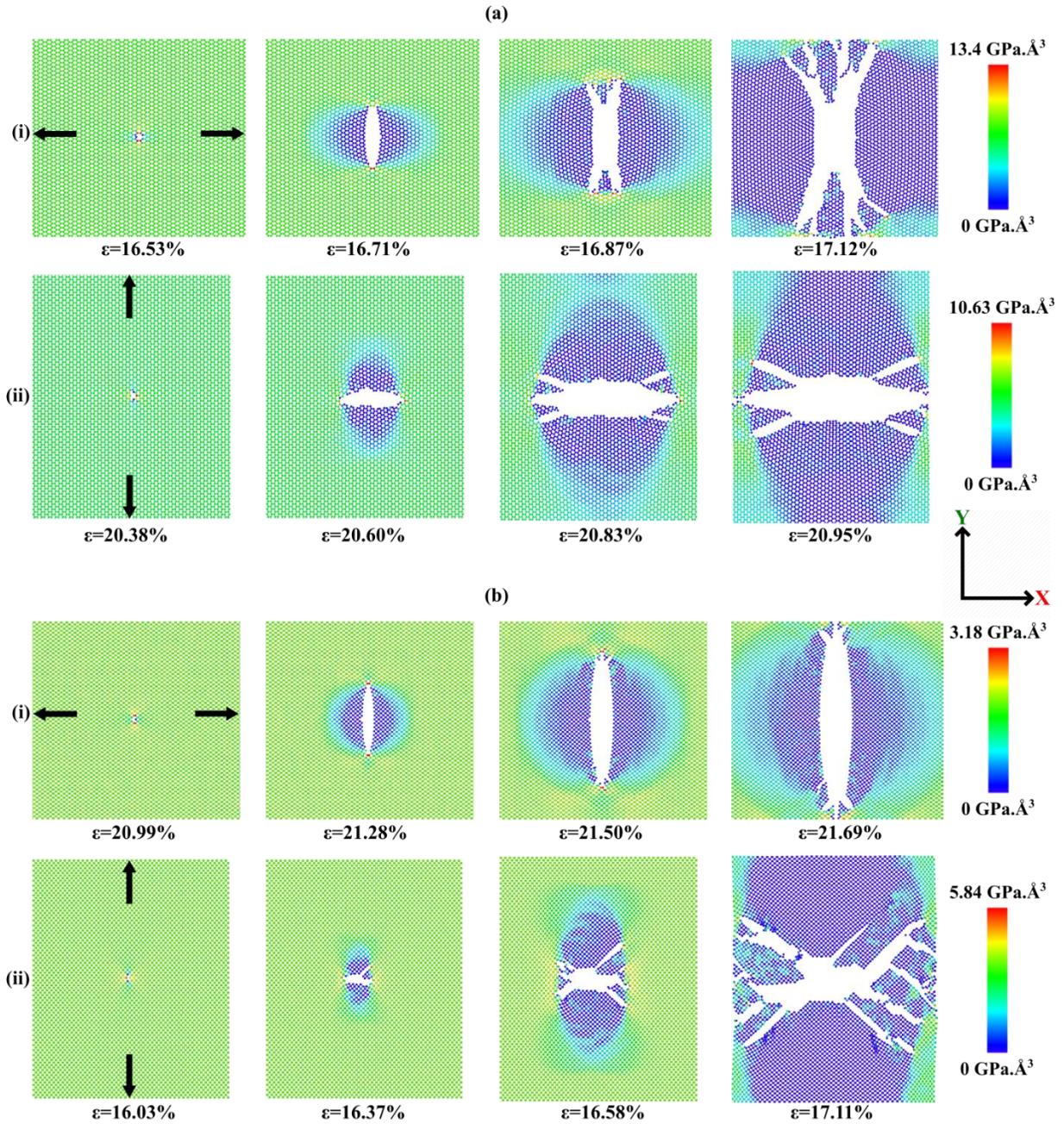

Figure 13: Per atom stress distribution and fracture process of monolayer (a) α (b) β structured Sb sheet having single vacancy at center, uniaxial tension (indicating by the arrow) along (i) armchair (ii) zigzag direction. X and Y axis denote armchair and zigzag edges respectively. The color bar signifies per atom stress in GPa.

## 5. CONCLUSIONS

For encapsulation, we conducted molecular dynamics simulations to explore the mechanical properties and fracture behavior of two distinct i.e. α and β structures of pristine and defected SLSb sheet. We examined the outcome of varying temperature, strain rate, crack length and defect percentage on the structural properties of the material. We showed that Sb allotropes can be modeled using LEFM with reasonable accuracy. Our calculations show that fracture toughness governs the material fracture for crack length ~60 Å and above, below which material strength governs the fracture. Increase in temperature and defect concentration not only deteriorate fracture strength and strain but also diminish the material soundness and elastic modulus. On the contrary, escalation in strain rate leads to the rise in fracture strength but Young's modulus remains the same. The vertical bond with XY plane in β-antimonene plays a vital role in determining the anisotropy of the material. And finally, fracture mechanism reveals that branching is prevailing in zigzag tension rather in armchair for both the structures.

## 7. REFERENCES

[1] P. Vogt, P. De Padova, C. Quaresima, J. Avila, E. Frantzeskakis, M.C. Asensio, A. Resta, B. Ealet, G. Le Lay, Silicene: Compelling Experimental Evidence for Graphenelike Two-Dimensional Silicon, Phys. Rev. Lett. 108 (2012) 155501. https://doi.org/10.1103/PhysRevLett.108.155501.
[2] S. Cahangirov, M. Topsakal, E. Aktürk, H. Şahin, S. Ciraci, Two- and One-Dimensional Honeycomb Structures of Silicon and Germanium, Phys. Rev. Lett. 102 (2009) 236804. https://doi.org/10.1103/PhysRevLett.102.236804.
[3] L. Song, L. Ci, H. Lu, P.B. Sorokin, C. Jin, J. Ni, A.G. Kvashnin, D.G. Kvashnin, J. Lou, B.I. Yakobson, P.M. Ajayan, Large scale growth and characterization of atomic hexagonal boron nitride layers, Nano Lett. 10 (2010) 3209–3215. https://doi.org/10.1021/nl1022139.
[4] A. Khandelwal, K. Mani, M.H. Karigerasi, I. Lahiri, Phosphorene – The two-dimensional black phosphorous: Properties, synthesis and applications, Mater. Sci. Eng. B. 221 (2017) 17–34. https://doi.org/10.1016/j.mseb.2017.03.011.
[5] H.-S. Tsai, S.-W. Wang, C.-H. Hsiao, C.-W. Chen, H. Ouyang, Y.-L. Chueh, H.-C. Kuo, J.-H. Liang, Direct Synthesis and Practical Bandgap Estimation of Multilayer Arsenene Nanoribbons, Chem. Mater. 28 (2016) 425–429. https://doi.org/10.1021/acs.chemmater.5b04949.
[6] J. Ji, X. Song, J. Liu, Z. Yan, C. Huo, S. Zhang, M. Su, L. Liao, W. Wang, Z. Ni, Y. Hao, H. Zeng, Two-dimensional antimonene single crystals grown by van der Waals epitaxy, Nat. Commun. 7 (2016) 1–9. https://doi.org/10.1038/ncomms13352.
[7] F. Reis, G. Li, L. Dudy, M. Bauernfeind, S. Glass, W. Hanke, R. Thomale, J. Schäfer, R. Claessen, Bismuthene on a SiC substrate: A candidate for a high-temperature quantum spin Hall material, Science. 357 (2017) 287–290. https://doi.org/10.1126/science.aai8142.
[8] Q. Ji, Y. Zhang, T. Gao, Y. Zhang, D. Ma, M. Liu, Y. Chen, X. Qiao, P.-H. Tan, M. Kan, J. Feng, Q. Sun, Z. Liu, Epitaxial Monolayer MoS2 on Mica with Novel Photoluminescence, Nano Lett. 13 (2013) 3870–3877. https://doi.org/10.1021/nl401938t.

[9] X. Lu, M.I.B. Utama, J. Lin, X. Gong, J. Zhang, Y. Zhao, S.T. Pantelides, J. Wang, Z. Dong, Z. Liu, W. Zhou, Q. Xiong, Large-Area Synthesis of Monolayer and Few-Layer MoSe2 Films on SiO2 Substrates, Nano Lett. 14 (2014) 2419–2425. https://doi.org/10.1021/nl5000906.

[10] C. Lan, C. Li, Y. Yin, Y. Liu, Large-area synthesis of monolayer WS2 and its ambient-sensitive photo-detecting performance, Nanoscale. 7 (2015) 5974–5980. https://doi.org/10.1039/C5NR01205H.

[11] M.M. Biener, J. Biener, C.M. Friend, Novel synthesis of two-dimensional TiS2 nanocrystallites on Au(111), J. Chem. Phys. 122 (2005) 034706. https://doi.org/10.1063/1.1826054.

[12] J. Zhou, J. Shi, Q. Zeng, Y. Chen, L. Niu, F. Liu, T. Yu, K. Suenaga, X. Liu, J. Lin, Z. Liu, InSe monolayer: synthesis, structure and ultra-high second-harmonic generation, 2D Mater. 5 (2018) 025019. https://doi.org/10.1088/2053-1583/aab390.

[13] Z. Ni, H. Bu, M. Zou, H. Yi, K. Bi, Y. Chen, Anisotropic mechanical properties of graphene sheets from molecular dynamics, Phys. B Condens. Matter. 405 (2010) 1301–1306. https://doi.org/10.1016/j.physb.2009.11.071.

[14] B. Uchoa, A.H. Castro Neto, Superconducting States of Pure and Doped Graphene, Phys. Rev. Lett. 98 (2007) 146801. https://doi.org/10.1103/PhysRevLett.98.146801.

[15] D. Jariwala, V.K. Sangwan, L.J. Lauhon, T.J. Marks, M.C. Hersam, Emerging Device Applications for Semiconducting Two-Dimensional Transition Metal Dichalcogenides, ACS Nano. 8 (2014) 1102–1120. https://doi.org/10.1021/nn500064s.

[16] D. Singh, S. K. Gupta, Y. Sonvane, I. Lukačević, Antimonene: a monolayer material for ultraviolet optical nanodevices, J. Mater. Chem. C. 4 (2016) 6386–6390. https://doi.org/10.1039/C6TC01913G.

[17] W. Choi, N. Choudhary, G.H. Han, J. Park, D. Akinwande, Y.H. Lee, Recent development of two-dimensional transition metal dichalcogenides and their applications, Mater. Today. 20 (2017) 116–130. https://doi.org/10.1016/j.mattod.2016.10.002.

[18] M. Pumera, Z. Sofer, 2D Monoelemental Arsenene, Antimonene, and Bismuthene: Beyond Black Phosphorus, Adv. Mater. 29 (2017) 1605299. https://doi.org/10.1002/adma.201605299.

[19] L. Li, Y. Yu, G.J. Ye, Q. Ge, X. Ou, H. Wu, D. Feng, X.H. Chen, Y. Zhang, Black phosphorus field-effect transistors, Nat. Nanotechnol. 9 (2014) 372–377. https://doi.org/10.1038/nnano.2014.35.

[20] O.Ü. Aktürk, V.O. Özçelik, S. Ciraci, Single-layer crystalline phases of antimony: Antimonenes, Phys. Rev. B. 91 (2015) 235446. https://doi.org/10.1103/PhysRevB.91.235446.

[21] S. Zhang, M. Xie, F. Li, Z. Yan, Y. Li, E. Kan, W. Liu, Z. Chen, H. Zeng, Semiconducting Group 15 Monolayers: A Broad Range of Band Gaps and High Carrier Mobilities, Angew. Chem. Int. Ed. 55 (2016) 1666–1669. https://doi.org/10.1002/anie.201507568.

[22] S. Zhang, Z. Yan, Y. Li, Z. Chen, H. Zeng, Atomically Thin Arsenene and Antimonene: Semimetal–Semiconductor and Indirect–Direct Band-Gap Transitions, Angew. Chem. Int. Ed. 54 (2015) 3112–3115. https://doi.org/10.1002/anie.201411246.

[23] Y. Wang, Y. Ding, Electronic Structure and Carrier Mobilities of Arsenene and Antimonene Nanoribbons: A First-Principle Study, Nanoscale Res. Lett. 10 (2015) 254. https://doi.org/10.1186/s11671-015-0955-7.

[24] G. Wang, R. Pandey, S.P. Karna, Atomically Thin Group V Elemental Films: Theoretical Investigations of Antimonene Allotropes, ACS Appl. Mater. Interfaces. 7 (2015) 11490–11496. https://doi.org/10.1021/acsami.5b02441.

[25] X. Wu, Y. Shao, H. Liu, Z. Feng, Y.-L. Wang, J.-T. Sun, C. Liu, J.-O. Wang, Z.-L. Liu, S.-Y. Zhu, Y.-Q. Wang, S.-X. Du, Y.-G. Shi, K. Ibrahim, H.-J. Gao, Epitaxial Growth and Air-Stability of Monolayer Antimonene on PdTe2, Adv. Mater. 29 (2017) 1605407. https://doi.org/10.1002/adma.201605407.

[26] P. Ares, F. Aguilar-Galindo, D. Rodríguez-San-Miguel, D.A. Aldave, S. Díaz-Tendero, M. Alcamí, F. Martín, J. Gómez-Herrero, F. Zamora, Mechanical Isolation of Highly Stable Antimonene under Ambient Conditions, Adv. Mater. 28 (2016) 6332–6336. https://doi.org/10.1002/adma.201602128.

[27] W. Yu, C.-Y. Niu, Z. Zhu, X. Cai, L. Zhang, S. Bai, R. Zhao, Y. Jia, Strain induced quantum spin Hall insulator in monolayer β-BiSb from first-principles study, RSC Adv. 7 (2017) 27816–27822. https://doi.org/10.1039/C7RA04153E.

[28] G. Pizzi, M. Gibertini, E. Dib, N. Marzari, G. Iannaccone, G. Fiori, Performance of arsenene and antimonene double-gate MOSFETs from first principles, Nat. Commun. 7 (2016) 1–9. https://doi.org/10.1038/ncomms12585.

[29] X. Chen, Q. Yang, R. Meng, J. Jiang, Q. Liang, C. Tan, X. Sun, The electronic and optical properties of novel germanene and antimonene heterostructures, J. Mater. Chem. C. 4 (2016) 5434–5441. https://doi.org/10.1039/C6TC01141A.

[30] Y.-Q. Wang, X. Wu, Y.-L. Wang, Y. Shao, T. Lei, J.-O. Wang, S.-Y. Zhu, H. Guo, L.-X. Zhao, G.-F. Chen, S. Nie, H.-M. Weng, K. Ibrahim, X. Dai, Z. Fang, H.-J. Gao, Spontaneous Formation of a Superconductor–Topological Insulator–Normal Metal Layered Heterostructure, Adv. Mater. 28 (2016) 5013–5017. https://doi.org/10.1002/adma.201600575.

[31] C.-H. Hsu, Z.-Q. Huang, F.-C. Chuang, C.-C. Kuo, Y.-T. Liu, H. Lin, A. Bansil, The nontrivial electronic structure of Bi/Sb honeycombs on SiC(0001), New J. Phys. 17 (2015) 025005. https://doi.org/10.1088/1367-2630/17/2/025005.

[32] S. Zhang, W. Zhou, Y. Ma, J. Ji, B. Cai, S.A. Yang, Z. Zhu, Z. Chen, H. Zeng, Antimonene Oxides: Emerging Tunable Direct Bandgap Semiconductor and Novel Topological Insulator, Nano Lett. 17 (2017) 3434–3440. https://doi.org/10.1021/acs.nanolett.7b00297.

[33] M. Fortin-Deschênes, O. Waller, T.O. Menteş, A. Locatelli, S. Mukherjee, F. Genuzio, P.L. Levesque, A. Hébert, R. Martel, O. Moutanabbir, Synthesis of Antimonene on Germanium, Nano Lett. 17 (2017) 4970–4975. https://doi.org/10.1021/acs.nanolett.7b02111.

[34] W. Lin, Y. Lian, G. Zeng, Y. Chen, Z. Wen, H. Yang, A fast synthetic strategy for high-quality atomically thin antimonene with ultrahigh sonication power, Nano Res. 11 (2018) 5968–5977. https://doi.org/10.1007/s12274-018-2110-0.

[35] Y. Shao, Z.-L. Liu, C. Cheng, X. Wu, H. Liu, C. Liu, J.-O. Wang, S.-Y. Zhu, Y.-Q. Wang, D.-X. Shi, K. Ibrahim, J.-T. Sun, Y.-L. Wang, H.-J. Gao, Epitaxial Growth of Flat Antimonene Monolayer: A New Honeycomb Analogue of Graphene, Nano Lett. 18 (2018) 2133–2139. https://doi.org/10.1021/acs.nanolett.8b00429.

[36] Y. Xu, B. Peng, H. Zhang, H. Shao, R. Zhang, H. Zhu, First-principle calculations of optical properties of monolayer arsenene and antimonene allotropes, Ann. Phys. 529 (2017) 1600152. https://doi.org/10.1002/andp.201600152.

[37] H.V. Phuc, N.N. Hieu, B.D. Hoi, L.T.T. Phuong, N.V. Hieu, C.V. Nguyen, Out-of-plane strain and electric field tunable electronic properties and Schottky contact of graphene/antimonene heterostructure, Superlattices Microstruct. 112 (2017) 554–560. https://doi.org/10.1016/j.spmi.2017.10.011.

[38] X. Fan, Y. Li, L. Su, K. Ma, J. Li, H. Zhang, Theoretical prediction of tunable electronic and magnetic properties of monolayer antimonene by vacancy and strain, Appl. Surf. Sci. 488 (2019) 98–106. https://doi.org/10.1016/j.apsusc.2019.05.133.

[39] S. Dai, Y.-L. Lu, P. Wu, Tuning electronic, magnetic and optical properties of Cr-doped antimonene via biaxial strain engineering, Appl. Surf. Sci. 463 (2019) 492–497. https://doi.org/10.1016/j.apsusc.2018.08.252.

[40] A. Bafekry, M. Ghergherehchi, S.F. Shayesteh, Tuning the electronic and magnetic properties of antimonene nanosheets via point defects and external fields: first-principles calculations, Phys. Chem. Chem. Phys. 21 (2019) 10552–10566. https://doi.org/10.1039/C9CP01378D.

[41] D.R. Kripalani, A.A. Kistanov, Y. Cai, M. Xue, K. Zhou, Strain engineering of antimonene by a first-principles study: Mechanical and electronic properties, Phys. Rev. B. 98 (2018) 085410. https://doi.org/10.1103/PhysRevB.98.085410.

[42] T. Rakib, S. Saha, M. Motalab, S. Mojumder, M.M. Islam, Atomistic Representation of Anomalies in the Failure Behaviour of Nanocrystalline Silicene, Sci. Rep. 7 (2017) 1–12. https://doi.org/10.1038/s41598-017-15146-6.

[43] MathWorks Announces Release 2018a of the MATLAB and Simulink Product Families, (n.d.). https://www.mathworks.com/company/newsroom/mathworks-announces-release-2018a-of-the-matlab-and-simulink-product-families.html (accessed July 4, 2019).

[44] P. Hirel, Atomsk: A tool for manipulating and converting atomic data files, Comput. Phys. Commun. 197 (2015). https://doi.org/10.1016/j.cpc.2015.07.012.

[45] J.-W. Jiang, Handbook of Stillinger-Weber Potential Parameters for Two-Dimensional Atomic Crystals, BoD – Books on Demand, 2017.

[46] A. Stukowski, Visualization and analysis of atomistic simulation data with OVITO–the Open Visualization Tool, Model. Simul. Mater. Sci. Eng. 18 (2009) 015012. https://doi.org/10.1088/0965-0393/18/1/015012.

[47] S. Plimpton, Fast Parallel Algorithms for Short-Range Molecular Dynamics, J. Comput. Phys. 117 (1995) 1–19. https://doi.org/10.1006/jcph.1995.1039.

[48] K.L. Grosse, M.-H. Bae, F. Lian, E. Pop, W.P. King, Nanoscale Joule heating, Peltier cooling and current crowding at graphene–metal contacts, Nat. Nanotechnol. 6 (2011) 287–290. https://doi.org/10.1038/nnano.2011.39.

[49] Q.-W. Wang, H.-B. Zhang, J. Liu, S. Zhao, X. Xie, L. Liu, R. Yang, N. Koratkar, Z.-Z. Yu, Multifunctional and Water-Resistant MXene-Decorated Polyester Textiles with Outstanding Electromagnetic Interference Shielding and Joule Heating Performances, Adv. Funct. Mater. 29 (2019) 1806819. https://doi.org/10.1002/adfm.201806819.

[50] M.A.N. Dewapriya, A.S. Phani, R.K.N.D. Rajapakse, Influence of temperature and free edges on the mechanical properties of graphene, Model. Simul. Mater. Sci. Eng. 21 (2013) 065017. https://doi.org/10.1088/0965-0393/21/6/065017.

[51] M. Motalab, R.A.S.I. Subad, A. Ahmed, P. Bose, R. Paul, J.C. Suhling, Atomistic Investigation on Mechanical Properties of Sn-Ag-Cu Based Nanocrystalline Solder Material, in: American Society of Mechanical Engineers Digital Collection, 2020. https://doi.org/10.1115/IMECE2019-12109.

[52] Q.-X. Pei, Z.-D. Sha, Y.-Y. Zhang, Y.-W. Zhang, Effects of temperature and strain rate on the mechanical properties of silicene, J. Appl. Phys. 115 (2014) 023519. https://doi.org/10.1063/1.4861736.

[53] Z. Yang, Y. Huang, F. Ma, Y. Sun, K. Xu, P.K. Chu, Temperature and strain-rate effects on the deformation behaviors of nano-crystalline graphene sheets, Eur. Phys. J. B. 88 (2015) 135. https://doi.org/10.1140/epjb/e2015-50850-x.

[54] W. Ding, D. Han, J. Zhang, X. Wang, Mechanical responses of WSe2 monolayers: a molecular dynamics study, Mater. Res. Express. 6 (2019) 085071. https://doi.org/10.1088/2053-1591/ab2085.

[55] A.P. Thompson, S.J. Plimpton, W. Mattson, General formulation of pressure and stress tensor for arbitrary many-body interaction potentials under periodic boundary conditions, J. Chem. Phys. 131 (2009) 154107. https://doi.org/10.1063/1.3245303.

[56] S.-H. Cheng, C.T. Sun, Size-Dependent Fracture Toughness of Nanoscale Structures: Crack-Tip Stress Approach in Molecular Dynamics, J. Nanomechanics Micromechanics. 4 (2014). https://doi.org/10.1061/(ASCE)NM.2153-5477.0000063.

[57] M.H. Rahman, S. Mitra, M. Motalab, P. Bose, Investigation on the mechanical properties and fracture phenomenon of silicon doped graphene by molecular dynamics simulation, RSC Adv. 10 (2020) 31318–31332. https://doi.org/10.1039/D0RA06085B.

[58] R.A.S.I. Subad, T.S. Akash, P. Bose, M.M. Islam, Engineered Defects to Modulate Fracture Strength of Single Layer MoS2: An Atomistic Study, ArXiv200103796 Cond-Mat. (2020). http://arxiv.org/abs/2001.03796 (accessed March 17, 2020).

[59] A.H.N. Shirazi, R. Abadi, M. Izadifar, N. Alajlan, T. Rabczuk, Mechanical responses of pristine and defective C3N nanosheets studied by molecular dynamics simulations, Comput. Mater. Sci. 147 (2018) 316–321. https://doi.org/10.1016/j.commatsci.2018.01.058.

[60] R. Ansari, S. Ajori, B. Motevalli, Mechanical properties of defective single-layered graphene sheets via molecular dynamics simulation, Superlattices Microstruct. 51 (2012) 274–289. https://doi.org/10.1016/j.spmi.2011.11.019.

[61] Z.-D. Sha, Q.-X. Pei, Y.-Y. Zhang, Y.-W. Zhang, Atomic vacancies significantly degrade the mechanical properties of phosphorene, Nanotechnology. 27 (2016) 315704. https://doi.org/10.1088/0957-4484/27/31/315704.

[62] P. Zhang, L. Ma, F. Fan, Z. Zeng, C. Peng, P.E. Loya, Z. Liu, Y. Gong, J. Zhang, X. Zhang, P.M. Ajayan, T. Zhu, J. Lou, Fracture toughness of graphene, Nat. Commun. 5 (2014) 1–7. https://doi.org/10.1038/ncomms4782.

**Supplementary Figures**

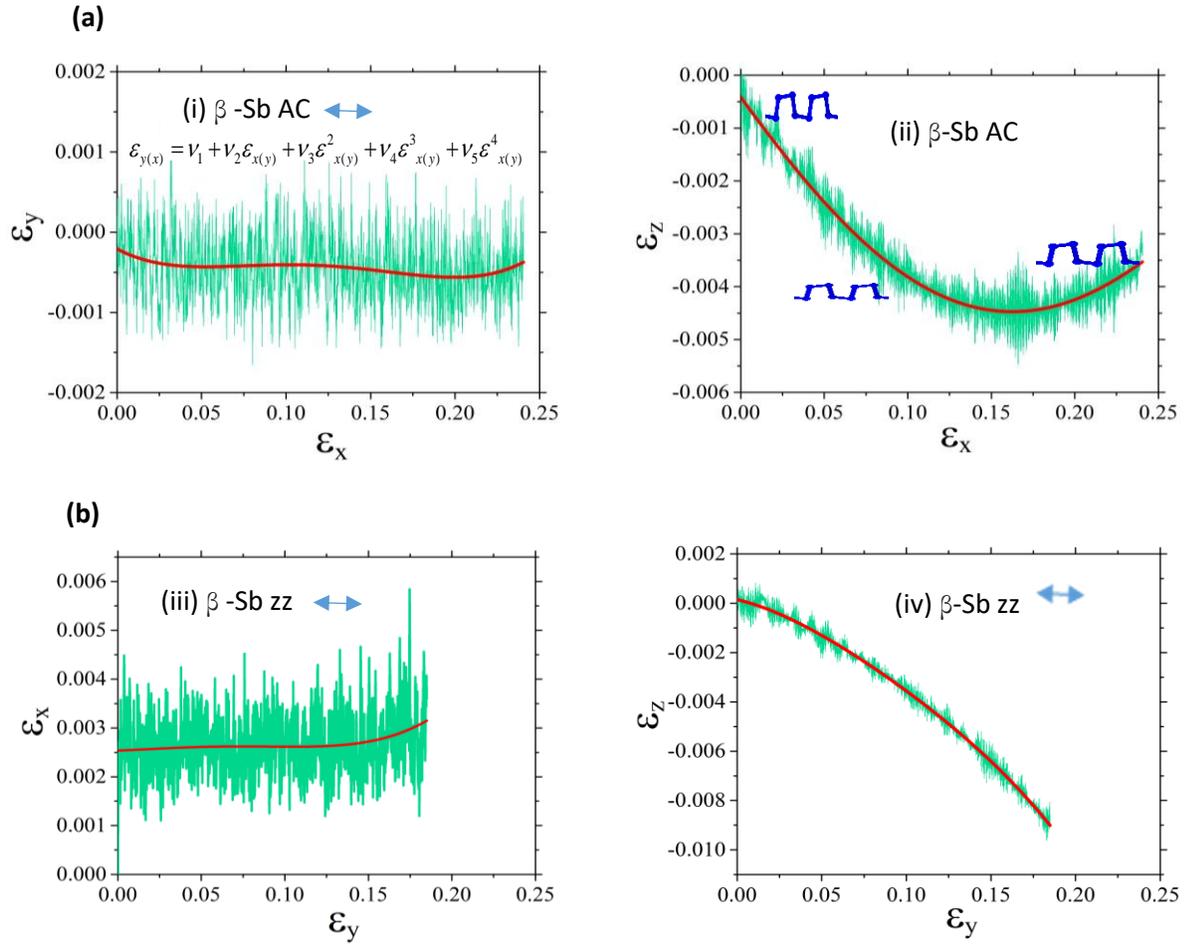

Figure S1: Strain responses of $\beta$-Sb to the non-loading axes for (a) armchair loading and (b) zigzag loading

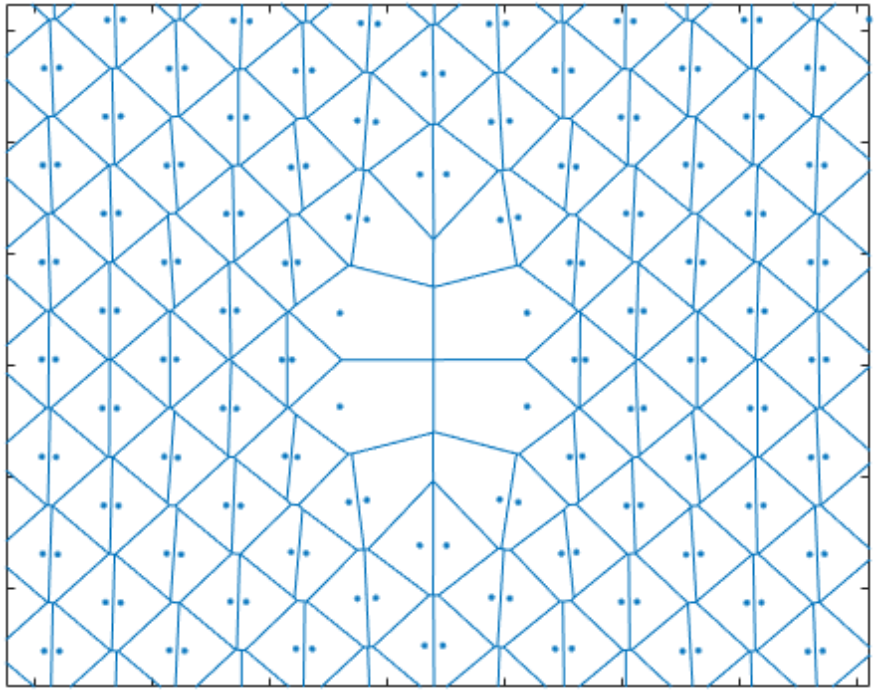

Figure S2: Voronoi areas for atoms in a cracked β-Sb sample. Volumes are calculated with a uniform thickness of 1 nm. Note that, volumes of crack edge atoms (except crack-tips) are ill-defined and not used in our calculations.